\documentclass[amsmath,amssymb,aps,twocolumn,pre,floatfix]{revtex4}

\usepackage{mathrsfs}
\usepackage{amsmath}
\usepackage{amssymb}
\usepackage{color}
\usepackage{graphicx}	
\usepackage{bm}
\usepackage{ulem} 
\usepackage{titlesec}
\titleformat{\paragraph}[runin]
{\bfseries\scshape}{\theparagraph}{1em}{}
\usepackage{braket}
\usepackage{multirow}

\newcommand{\be}{\begin{equation}}
\newcommand{\ee}{\end{equation}}
\newcommand{\bef}{\begin{figure}}
\newcommand{\eef}{\end{figure}}
\newcommand{\bea}{\begin{eqnarray}}
\newcommand{\eea}{\end{eqnarray}}

\begin{document}
\title{Biohybrid membrane formation by directed insertion of Aquaporin in to a solid-state nanopore}
%
\author{Fran\c cois Sicard}
\author{A. Ozgur Yazaydin}
\thanks{\texttt{ozgur.yazaydin@ucl.ac.uk}.}
\affiliation{Department of Chemical Engineering, University College London, WC1E 7JE London, UK}

\begin{abstract}
%
%
Technical challenges in molecule sensing and chemical detection 
have created an increasing demand for transformative materials with 
high sensitivity and specificity.
Biohybrid nanopores have attracted growing interest as 
they can ideally combine the durability of solid-state nanopores 
with the precise structure of biological nanopores. 
Particular care must be taken to control how biological nanopores adapt 
to their surroundings once in contact with the solid-state nanopore.
%
Two major challenges are to precisely control this adaptability under 
dynamic conditions and provide predesigned functionalities that can be 
manipulated 
for engineering applications.
Here, we report on the computational design of a distinctive class 
of biohybrid active membrane layer, built from the directed-insertion of 
aquaporin-incorporated lipid shell in to a silica nanopore.
%
First, we describe in detail the mechanisms at play in the insertion 
of the 
biological membrane 
in to the solid-state nanopore.
Then we analyze the 
structural stability of the system and demonstrate that its water permeability is comparable 
to the one measured in the biological environment.
Finally, we discuss how the technology implemented could 
be applicable to 
environmental and biomedical applications, such as water 
desalination and drug discovery,
where targeting and controlled permeation 
of small molecules must be efficiently addressed.
\end{abstract}
\date{\today}

\maketitle

Considerable attention has been devoted to the design, 
characterization and development of biohybrid nanopore based platforms 
due to their potential application in the areas of environment and healthcare including molecule sensing~\cite{2019-NatComm-Cao-DalPeraro,2020-Nanoscale-Nicolai-Senet}, disease diagnosis~\cite{2020-FrontGenet-Minervini-Albano}, 
drug design~\cite{2020-ChemSci-Wang-Edel}, chemical detection~\cite{2019-ACSNano-Shoji-White}, 
pollutant removal~\cite{2015-DWT-Esmaeilian-Liaghat}, and water desalination~\cite{2018-ECL-Ramanathan-Rawajfeh}. 
In particular, biohybrid membranes have received growing interest 
over the last decade as they can ideally combine the durability 
of solid-state nanopores with the precise structure and functionality of biological nanopores~\cite{2013-NanoToday-Haque-Guo}. 
In this perspective, tremendous progress has been made regarding 
the incorporation of biological compounds inside synthetic pores by modifying their surface chemistry. This includes, 
among others, the coating of nanopores 
with supported fluid lipid bilayer~\cite{2011-NatureNano-Yusko-Mayer,2013-CSBB-Sun-Armugam,2013-JMS-Wang-Hong,2015-JMCA-Ding-Gao,2019-Desalination-Fuwad-Jeon} and the insertion of 
biological structures in silicon based substrates such as 
the $\alpha$-haemolysin protein~\cite{2010-NatureNano-Hall-Dekker}, DNA origami structures~\cite{2012-NanoLett-Bell-Keyser}, 
and biological nanodiscs~\cite{2018-Nanomaterials-Farajollahi-Gliemann}.\\

We report here on the computational design of a distinctive class of 
biohybrid active membrane layer 
built from the directed insertion of aquaporin-incorporated 
lipid shell in a model silica pore, which could pave the way to 
the development of transformative devices for water filtration and drug discovery. 
Aquaporins (Aqp) are transmembrane proteins 
that are present in the biological membranes of organisms such as plants, 
mammals and bacteria~\cite{2013-CurrBiol-Verkman}. They form pores that act as water channels
allowing water molecules 
to move across the plasma membrane while rejecting protons, charged 
particles and other solutes.
Of particular interest 
is the bacterial Aquaporin Z (AqpZ), 
expressed in \textit{Escherichia coli}, which is the smallest, simplest 
and most robust member in the Aqp family~\cite{2000-MolMicrobiol-Calamita}. 
Along with its high selectivity and water permeability, AqpZ 
has been purified in 
high concentration allowing it to be used for integration into biomimetic 
membranes for water filtration~\cite{2013-CSBB-Sun-Armugam,2016-PBiochem-Hang-Xu}.\\

\begin{figure}[t]
\includegraphics[width=0.9 \columnwidth, angle=-0]{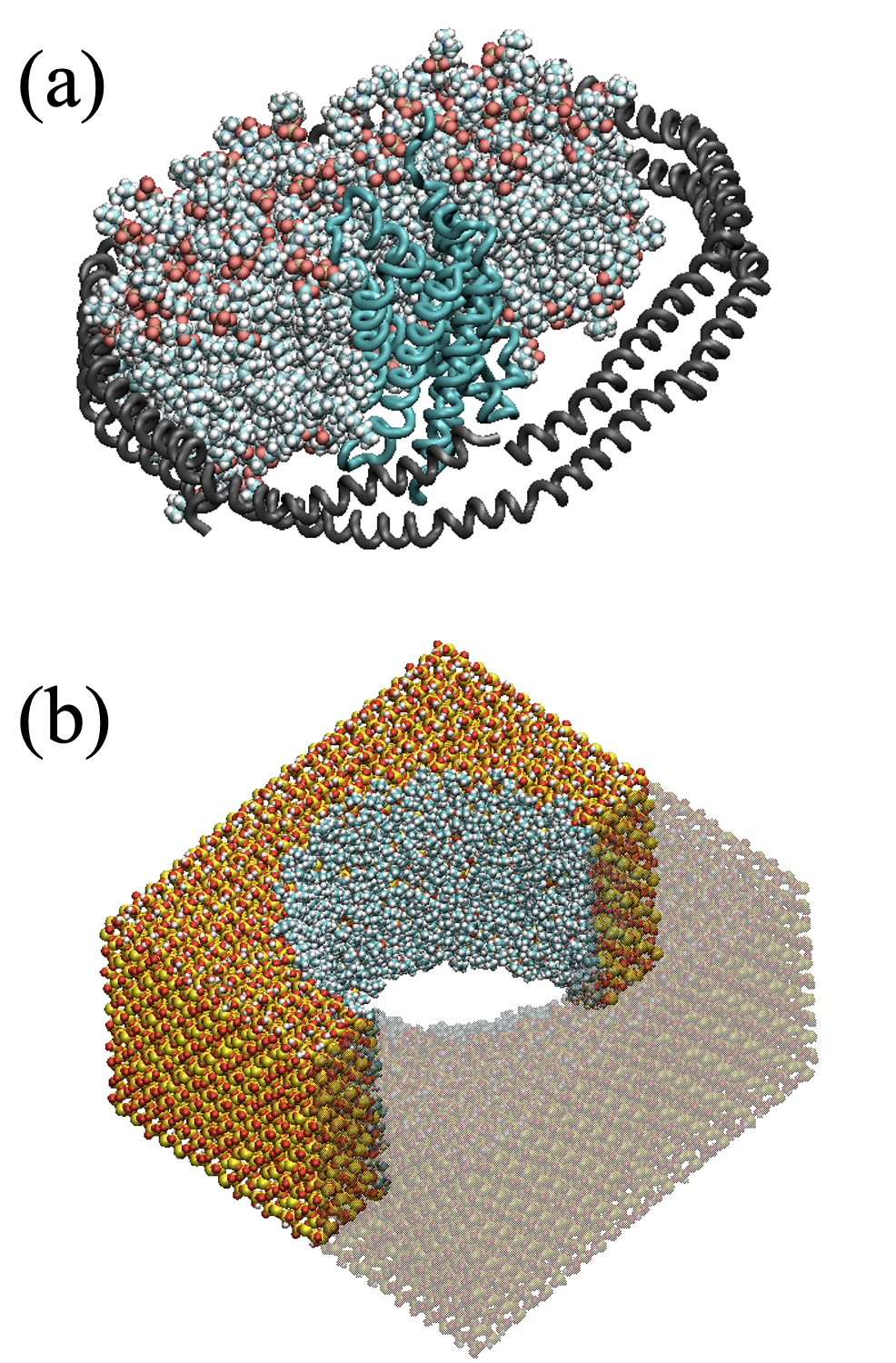}
 \caption{Schematic representation of (a) the 
 AqpZ-incorporated lipid nanodisc stabilized by the MSP1E2D1 membrane scaffold protein and (b) the cylindrical 
 nanopore carved out of a silica-cristobalite slab. The external and internal surface of the slab is 
 functionalized with hydroxyl and propyl groups,
 respectively. Half of the 
 nanopore is made transparent and half of the lipid 
 bilayer is omitted for clarity.
 The backbone of the AqpZ monomer and the MSP 
 are shown in cyan and dark grey, respectively. 
 Yellow, red, blue, white, gold, and cyan spheres 
 represent silicon, oxygen, nitrogen, hydrogen, 
 phosphate and carbon atoms, respectively.}
\label{fig1}
 \end{figure}

In this work, 
AqpZ is originally embedded in a lipid nanodisc stabilized by 
membrane scaffold proteins (MSPs), as shown in Fig.~\ref{fig1}. MSPs are charged 
helical-amphiphatic proteins that wrap around the lipid bilayers 
with precisely defined inner and outer diameters~\cite{2009-ME-Ritchie-Sligar}. MSP-based lipid nanodiscs 
are known to provide a native membrane environment that aids 
in preparing integral membrane proteins in biologically active folded 
forms for structural studies~\cite{2016-NSMB-Denisov-Sligar}.
They have recently been shown to serve as  biological template 
that can be electrochemically inserted into solid-state nanopore~\cite{2018-Nanomaterials-Farajollahi-Gliemann}. 
This integration could be a promising step towards 
reproducible mass production of biohybrid membranes. 
However, it still faces structural limitations associated with the positioning of the MSP-based 
nanodisc and its interaction with the solid-state nanopore, 
which affects the sealing performance of the 
membrane
and limits its application in 
biohybrid nanopore technology.
 
%
%

To circumvent this limitation, 
we show how the AqpZ-incorporated lipid 
nanodisc can be dissociated from its 
surrounding MSPs 
and inserted into functionalized silica 
nanopore to form a functional biohybrid membrane layer.
%
%
We first employ non-equilibrium molecular dynamics 
(NEMD) simulation to study   
the dissociation of the AqpZ-incorporated lipid nanodisc  
from the immobilised MSPs under the application 
of an external pressure gradient and its subsequent insertion 
into the pore. 
In this process, we show that the thickness  
of the 
solid-state nanopore and the nature of the functionalizing groups
must be properly chosen with respect to the
characteristics of the MSP-based lipid nanodisc to ensure the final biohybrid nanopore is stable and it is leak proof at the interface of the functionalized nanopore and the lipid shell.
In particular, the preferential interaction between the hydrocarbon tail of the lipid molecules and the interior surface of the pore 
yields the expulsion of 
the water molecules and therefore is responsible for tuning 
the sealing performance of the membrane.
We then investigate the structural stability 
of the biohybrid membrane layer 
in pure and saline water conditions with equilibrium MD simulations.
Finally, we study the non-equilibrium osmotic permeability of the system when an external pressure gradient is applied 
and compare the results with those measured in biological environments.


\section*{Results and Discussion}
\textbf{System Characteristics.} We first give the 
characteristics of the MSP-based lipid nanodisc and 
solid-state nanopore considered in this work. 
A schematic representation of the system is shown 
in Fig.~\ref{fig1}.
One AqpZ monomer is originally embedded in a 
$1$-palmitoyl-$2$-oleoyl-sn-glycero-$3$-phosphocholine 
(POPC) lipid-nanodisc stabilized by the MSP1E2D1 membrane 
scaffold protein. The diameter of the corresponding nanodisc 
is $11.1$ nm~\cite{2021-Membranes-Errasti-Palacin}.
A cylindrical nanopore is carved out of a silica-cristobalite 
slab which surface was functionalized with hydroxyl 
groups (-OH). The interior of the pore is functionalized 
with propyl groups ($\textrm{-C}_3\textrm{H}_7$) to render 
its surface hydrophybic. The length of the functionalizing group 
is specifically chosen such that it confers adequate sealing 
properties of the nanopore due to preferential interaction 
between the hydrocarbon tails of the lipid shell and 
the functionalizing groups. The comparison with 
shorter groups is shown in the 
Supporting Information (SI). 
The height and diameter of the cylindrical nanopore 
were specifically chosen to allow the insertion 
of the AqpZ-incorporated lipid shell, while preventing 
the insertion of the MSPs.
%
The CHARMM force field~\cite{2016-NatureMethods-Huang-MacKerell} was 
employed to model the biological entities 
(MSPs, AqpZ, POPC) and the propyl group on 
the silica surface. 
The INTERFACE force field~\cite{2013-Langmuir-Heinz-Emami} and TIP3P potential~\cite{1983-JChemPhys-Jorgensen-Madura} were employed 
to model the silica nanopore and water molecules, respectively, 
as they are compatible with the CHARMM force field.
The simulations were conducted using 
the GROMACS software package, 
version 2018.8~\cite{2015-SoftwareX-Abraham-Lindahl}. 
The energy of the system was first optimized using 
the steepest descent minimization algorithm.
The system was then equilibrated 
at constant temperature, $T=300$ K, finishing with equilibration 
at constant pressure, $P=1$ bar, to equilibrate the 
fluid density. 
Finally, a production simulation was carried out in the  
NVT ensemble. 
%
NEMD simulations were performed using the extension 
of the pull module of the GROMACS software 
developed by Gr{\"a}ter and coworkers~\cite{2019-BJ-Herrera-Grater}.
The details of the numerical parameterization and the system equilibration 
are given in 
the SI.\\

\begin{figure*}[t]
\includegraphics[width=1.0 \textwidth, angle=-0]{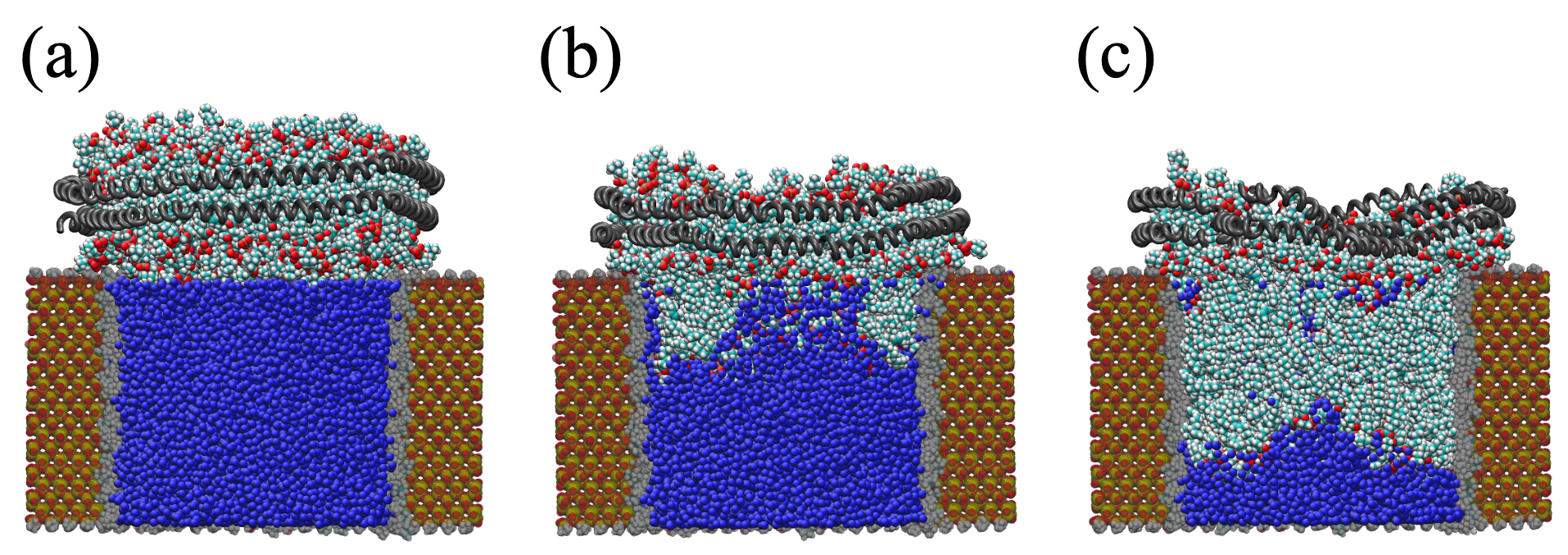}
 \caption{Schematic representation of the insertion 
 of the AqpZ-incorporated lipid nanodisc into 
 the solid-state nanopore. (a) Under the 
 application of the hydrostatic pressure 
 difference, the lipid nanodisc first comes 
 in contact with the silica slab. 
 (b) As the system moves further down 
 the pore, the outer edge of the lipid nanodisc 
 acts as a support layer that stabilizes 
 the membrane scaffold proteins (MSPs) and prevents 
 them to penetrate inside the cavity. 
 (c) The hydrophobic tails of the lipid molecules  interact favorably with the propyl groups 
 functionalizing the interior surface of the 
 pore, which yields the depletion of the water 
 molecules.
 The backbone of the MSPs is shown in dark grey. 
 Yellow, red, blue, white, gold, and cyan spheres 
 represent silicon, oxygen, nitrogen, hydrogen, 
 phosphate and carbon atoms in the lipid molecules 
  and silica surface, respectively. The oxygen 
  atoms of the water molecules are shown in blue. 
  Only half of the pore and the water molecules in the pore are shown for clarity.}
\label{fig2}
 \end{figure*}

\textbf{Directed insertion of AqpZ-incorporated lipid nanodisc in to a solid-state nanopore.} 
The MSP-based lipid nanodisc is initially positioned above the 
cylindrical pore, as shown in Fig.~\ref{fig2}. 
Farajollahi \textit{et al.} have recently demonstrated that 
the positioning of MSP-based nanodiscs in solid 
state nanopores could be effectively driven by 
applying concentration gradient-based or electrophoresis-driven diffusion~\cite{2018-Nanomaterials-Farajollahi-Gliemann}. 
A water pressure gradient is applied across the 
membrane to dissociate the AqpZ-incorporated lipid 
shell from the MSPs and to direct its  
insertion in the cylindrical pore. 
Using the numerical method of 
Zhu \textit{et al.}~\cite{2002-BJ-Zhu-Schulten,2004-BJ-Zhu-Schulten},
a hydrostatic pressure difference 
of $\Delta P \sim 80$ MPa is created across the pore
 and is applied for $10$ ns, while the position of the solid-state membrane was restrained.
The value of $\Delta P$ was  
chosen to achieve the insertion of the 
nanodisc in a reasonably short simulation 
time.
For comparison, the free energy profile 
associated with the equilibrium dissociation of the AqpZ monomer from the MSP-based 
lipid nanodisc is shown in the SI.

Under the application of the hydrostatic pressure difference, 
the lipid nanodisc first comes  in contact with the silica slab. Due to preferential interaction between the hydrophilic head of the lipid molecules and the hydroxyl group on the surface of the slab, 
the outer edge of the lipid nanodisc acts as a support layer that stabilizes the MSPs and prevents them
 to penetrate inside the cavity (see Fig~\ref{fig2}a). 
As the AqpZ-incorporated lipid shell penetrates 
further in the solid-state membrane, the hydrophobic 
tails of the lipid molecules 
start interacting favorably with the propyl 
groups. This preferential interaction yields 
the depletion of the water molecules in the pore.
The depletion of water 
does not evenly happen around the interior surface 
of the pore at first, as water molecules 
preferentially interact with the hydrophilic head 
of the lipid molecules (see Fig~\ref{fig2}b).
As the simulation progresses,  
the AQP-incorporated lipid shell gradually moves 
further down the pore under the effect of the hydrostatic 
pressure difference. As a result, the inserted lipid 
molecules evenly spreads in the solid-state membrane,
leading to the complete depletion of water around 
the Aqp-incorporated lipid shell 
(see Fig~\ref{fig2}c). The water pressure difference 
across the membrane is then switched off and 
the system is relaxed in a $\sim 200$ ns simulation.\\

\begin{figure}[t]
\includegraphics[width=1.0 \columnwidth, angle=-0]{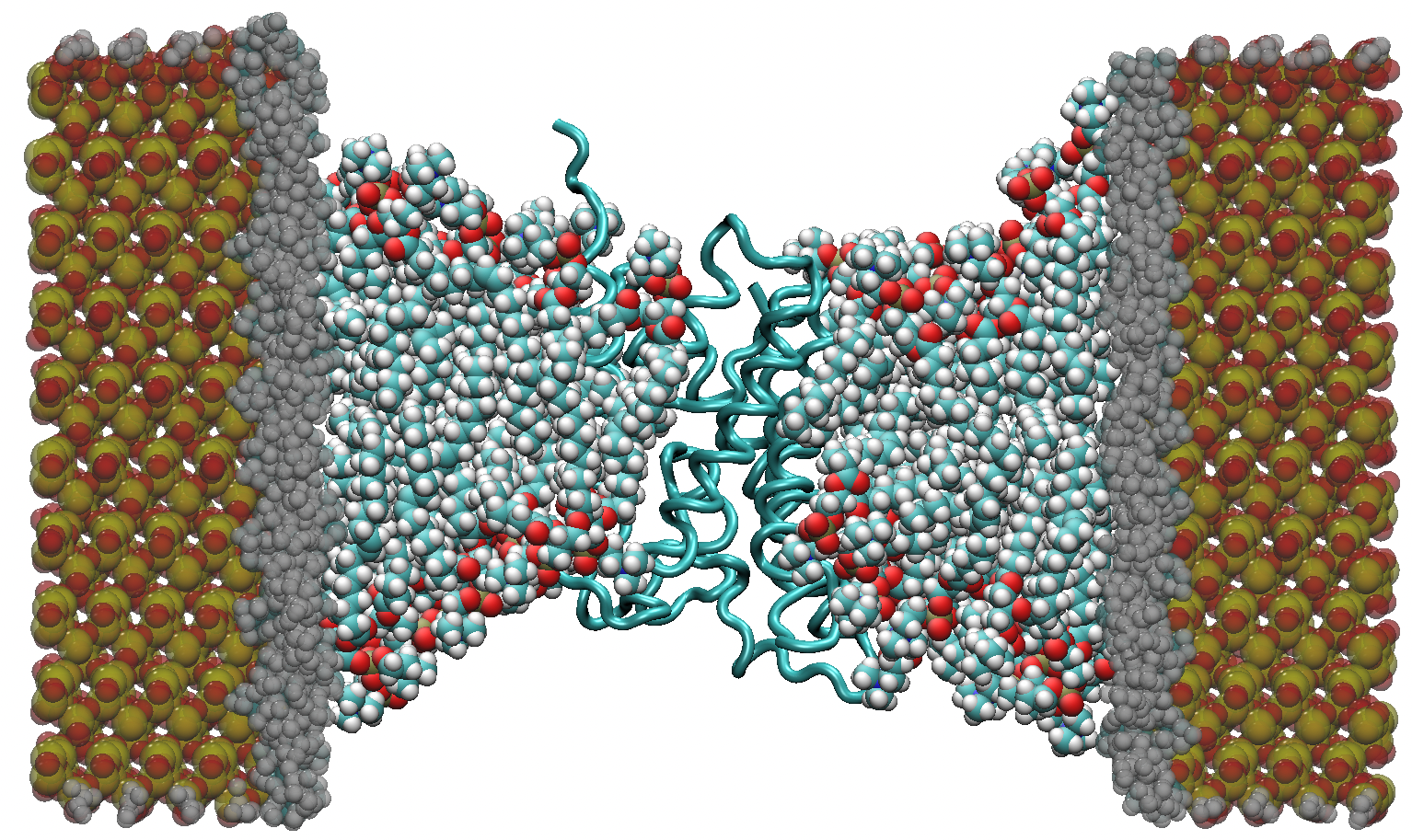}
 \caption{Schematic representation of the 
 Aqp-incorporated lipid shell inserted in the 
 silica nanopore after the system relaxed 
 for $200$ ns. The backbone of the AqpZ monomer 
 is shown in cyan. Yellow, red, blue, 
 white, gold, and cyan spheres 
 represent silicon, oxygen, nitrogen, hydrogen, 
 phosphate and carbon atoms, respectively. 
 The water molecules are not shown and the 
 atoms composing the functionalized silica 
 pore are shadowed for clarity. We observe 
 the progressive increase of the thickness 
 of the bilayer as the lipid molecules 
 radially position away from the center of 
 the protein.}
\label{fig3}
 \end{figure}

\begin{figure}[b]
\includegraphics[width=1.0 \columnwidth, angle=-0]{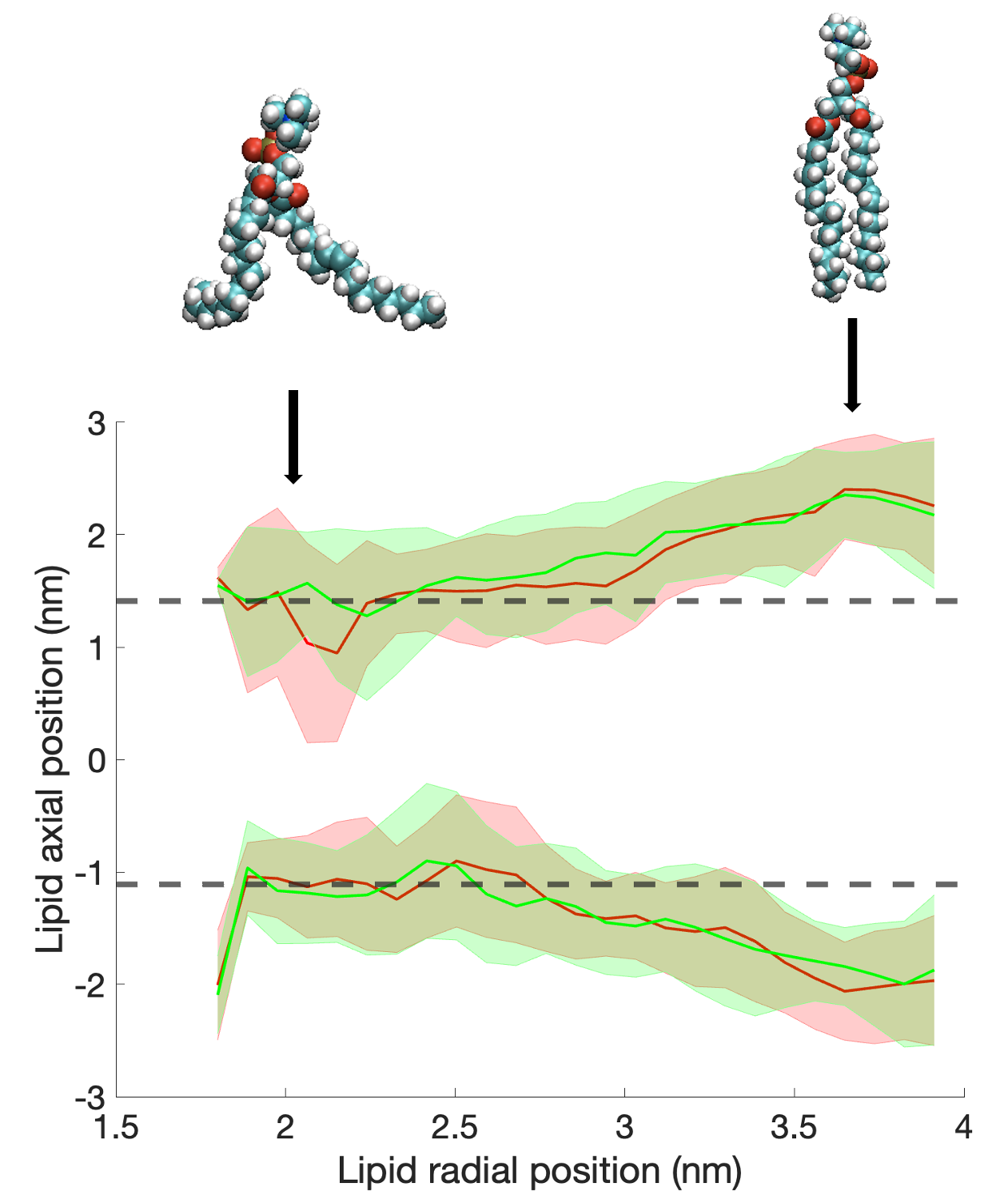}
 \caption{Distribution of the lipid 
 molecules within the ApqZ-incorporated lipid 
shell inserted into the solid-state nanopore, as 
a function of their radial and axial position.
The mean value of the position distribution 
of the lipid molecules in the biological 
environment is shown for comparison. Both the 
positions of the upper and lower leaflets are 
shown in pure (green) and saline (red) water.
The thickness of the bilayer progressively 
increases as the lipid molecules radially position 
away from the center of the pore. 
The AqpZ monomer and the interior surface 
of the pore are positioned at $\rho <2$ nm and 
$\rho \sim 4$ nm, respectively. 
Uncertainties, defined as the standard error, 
are represented by the shaded area.}
\label{fig4-draft}
 \end{figure}

\textbf{Stability of the active layer 
in pure water.} 
We first study the structural stability of the 
biohybrid membrane layer in pure water, 
\textit{i.e.} in the absence of ions.
%
%
A representative snapshot of the system after 
relaxation is shown in Fig.~\ref{fig3}. 
We observe the increase of the thickness of 
the bilayer as the lipid molecules radially position 
away from the center of the protein.
This evolution is quantitatively assessed in 
Fig.~\ref{fig4-draft}, where the axial position 
of the lipid molecules 
is shown as a function of their radial 
position with respect to the center of the pore. 
The average value of the position distribution 
of the lipid molecules measured in the biological environment, corresponding to a membrane 
thickness $h_{\textrm{bio}} \sim 2.5$ nm, 
is shown for comparison.
As the lipid molecules position closer to 
the surface of the pore, the thickness of 
the bilayer progressively increases, 
departing from the thickness measured in the 
biological environment.
This structural deformation is associated with 
the configurational change of the hydrocarbon 
chain of the lipid molecules with a transition 
from a Gauche conformation near the center of 
the nanopore to a nearly all-trans conformation 
at the pore surface, as shown in Fig.~\ref{fig4-draft}. 
This transition is due to preferential 
Van der Waals interaction between 
 the hydrophobic tail of lipids and 
the hydrophobic surface functionalized with the 
propyl groups~\cite{2018-Langmuir-Sicard-Striolo,2021-MSDE-Sicard-Striolo}. \\

\begin{figure}[t]
\includegraphics[width=1.0 \columnwidth, angle=-0]{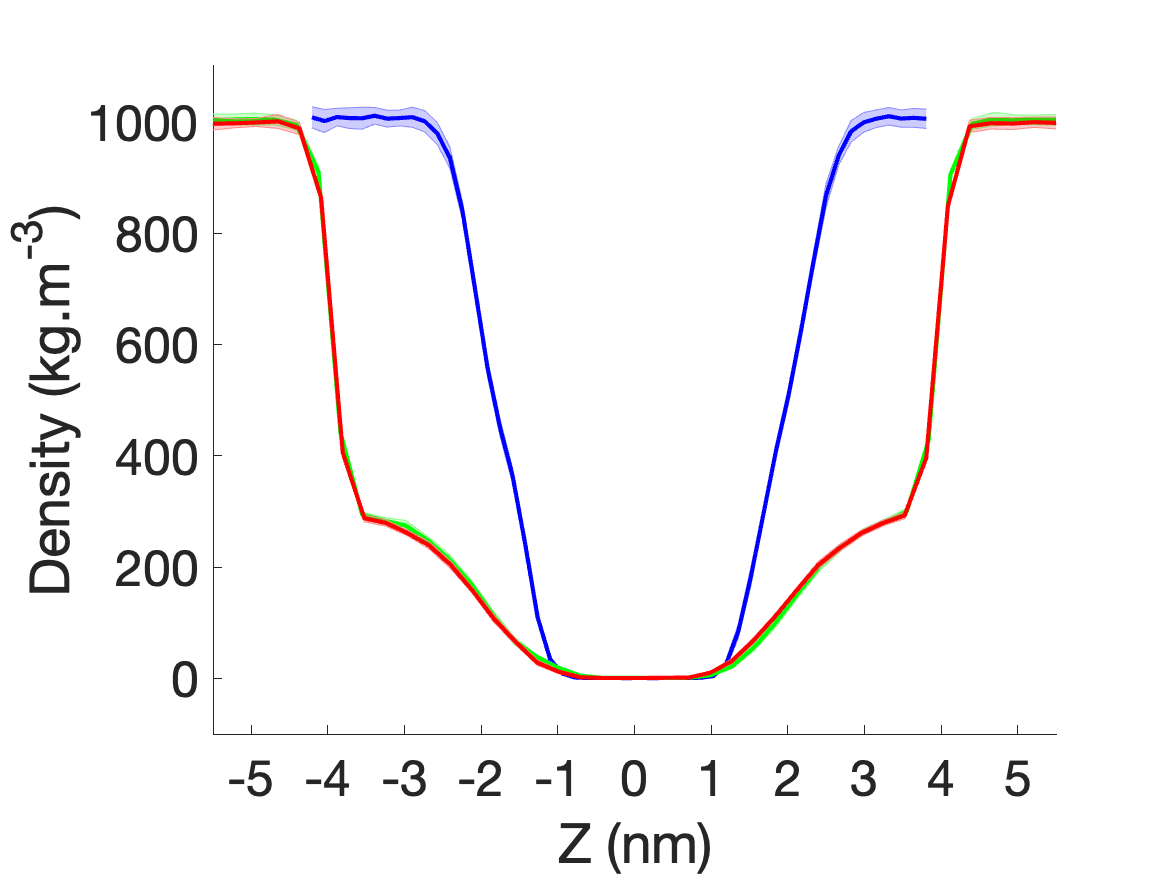}
 \caption{Density profile of the water 
 molecules along the Z direction of the 
 simulation box and centered around the 
 ApqZ monomer. Following the directed insertion 
 of the Aqp-incorporated nanodisc into the 
 solid-state membrane, we let the system relax
 for $200$ ns in fresh (green) and 
 sea (red) water. The density profile 
 of the water molecules in the biological 
 environment is shown for comparison (blue). 
 Uncertainties, defined as the standard error, 
are represented by the shaded area.}
\label{fig5-draft}
 \end{figure}

To further assess the stability of the biohybrid 
membrane and its ability to serve as a separator, 
we measured in Fig.~\ref{fig5-draft} the density distribution of the water molecules centered around the AqpZ monomer in the simulation box after relaxation
The density distribution of water measured 
in the biological environment is shown 
for comparison. 
For $\vert z \vert \leq 1$,  the profile of the water density is similar to 
the one observed in the biological simulation. 
For $1 < \vert z \vert \leq 4$ we observe 
the progressive increase of the water density, 
which happens at a slower rate than in 
the biological environment. 
This difference is explained by the 
conformational organisation of the 
lipid bilayer in the nanopore due to the preferential 
interaction with the hydrophobic surface, as shown 
in Fig.~\ref{fig3}. 
For $\vert z \vert > 4$, \textit{i.e.} outside 
the solid-state nanopore, the system transitions 
abruptly to reach the bulk water density similar 
to the one measured in the biological environment.\\

\textbf{Stability of the active layer 
in sea saline water.}  
We extend the analysis above to the study of 
the structural stability of the active membrane 
layer in the presence of chloride and sodium ions 
at characteristic salinity of sea water. 
In this condition, about $3.5\%$ of the weight of seawater comes from the dissolved 
salts~\cite{2016-PNAS-Cullum-Joshi}. 
%
The evolution of the thickness of the biological 
membrane as a function of the radial position 
of the lipid molecules is shown in Fig.~\ref{fig4-draft}. We observe the 
increase of the thickness of the bilayer, similar 
to the evolution obtained in pure water, 
as the lipid molecules radially position away from the center 
of the protein.
In addition, we show in Fig.~\ref{fig5-draft} the density distribution of water molecules in 
the simulation box. 
%
The density profile is similar to the one 
measured in pure water. This indicates that 
the presence of ions does not affect the structural 
stability of the Aqp-incorporated lipid shell 
once inserted in the solid-state nanopore.
Furthermore, we studied the density 
distribution of ions 
in the simulation box after relaxation. 
The results are shown in Fig.~S3a in the 
SI. Visual inspection shows that the ions do 
not penetrate the water channel of AqpZ or 
the lipid shell.
This is quantitatively assessed in Fig.~S3b in the SI, where we measured  the density distribution of ions in the 
simulation box. In particular, the ion density 
for $\vert z \vert \leq 1$ is 
equal to zero. For $1 < \vert z \vert \leq 4$ 
we observe the progressive increase of the ion density due to the increase of the thickness 
of the biological membrane, similar to what 
we obtained for the density distribution 
of the water molecule. 
Finally, for $\vert z \vert > 4$, the system transitions 
abruptly to reach the bulk ion density.\\

\textbf{Water transport across the active membrane layer.} 
We complete the analysis above 
with the study of the osmotic permeability of the 
biohybrid active membrane layer.
In the condition of reverse osmosis, the ability of 
the system to conduct water is characterized by 
the ratio of the net water flow to the hydrostatic pressure 
difference, $\Delta P$, across the membrane.
%
%
The volume flux, $J_v$, defined as 
the net flow of water per 
unit area of the membrane is obtained as
\begin{equation}
    J_v = L_P \Delta P
\label{hydraulicPerm}
\end{equation}
with $L_P$ the hydraulic permeability 
of the membrane. 
For comparaison, when the two sides of the membrane 
have the same hydrostatic pressure but different 
concentrations of an impermeable solute, an osmotic 
pressure difference is established and water will flow 
from the side with lower solute concentration to the 
other side. In dilute solutions, the molar water flux, 
$J_m$, is linearly proportional to the solute 
concentration difference $\Delta C$
\begin{equation}
    J_m = P_f \Delta C
\end{equation}
with $P_f$ the osmotic permeability of the membrane. 
The water flux generated due to a solute concentration 
difference is identical to that of generated 
by a hydrostatic pressure difference 
$\Delta P = R T \Delta C$, with $R$  the gas constant 
and $T$ the temperature of the system. 
Therefore, $L_P$ and $P_f$ are related by a constant 
factor
\begin{equation}
    P_f = (RT/V_W) L_P \,,
\label{OsmoticHydraulic}
\end{equation}
with $V_W$ the molar volume of water.\\

To determine the hydraulic permeability of the 
active membrane layer, a water pressure gradient is 
applied along the direction orthogonal to the 
membrane. 
The position of the membrane is restrained along 
the direction of the pressure gradient to avoid 
undesirable translation of the system. 
Details of the numerical implementation is given 
in the SI.
The water flux, $J_W$, through the channel in 
the membrane can then be measured at the 
atomistic level by counting the net number of water molecules 
passing through the channel during the simulation.  

\begin{figure}[b]
\includegraphics[width=1.0 \columnwidth, angle=-0]{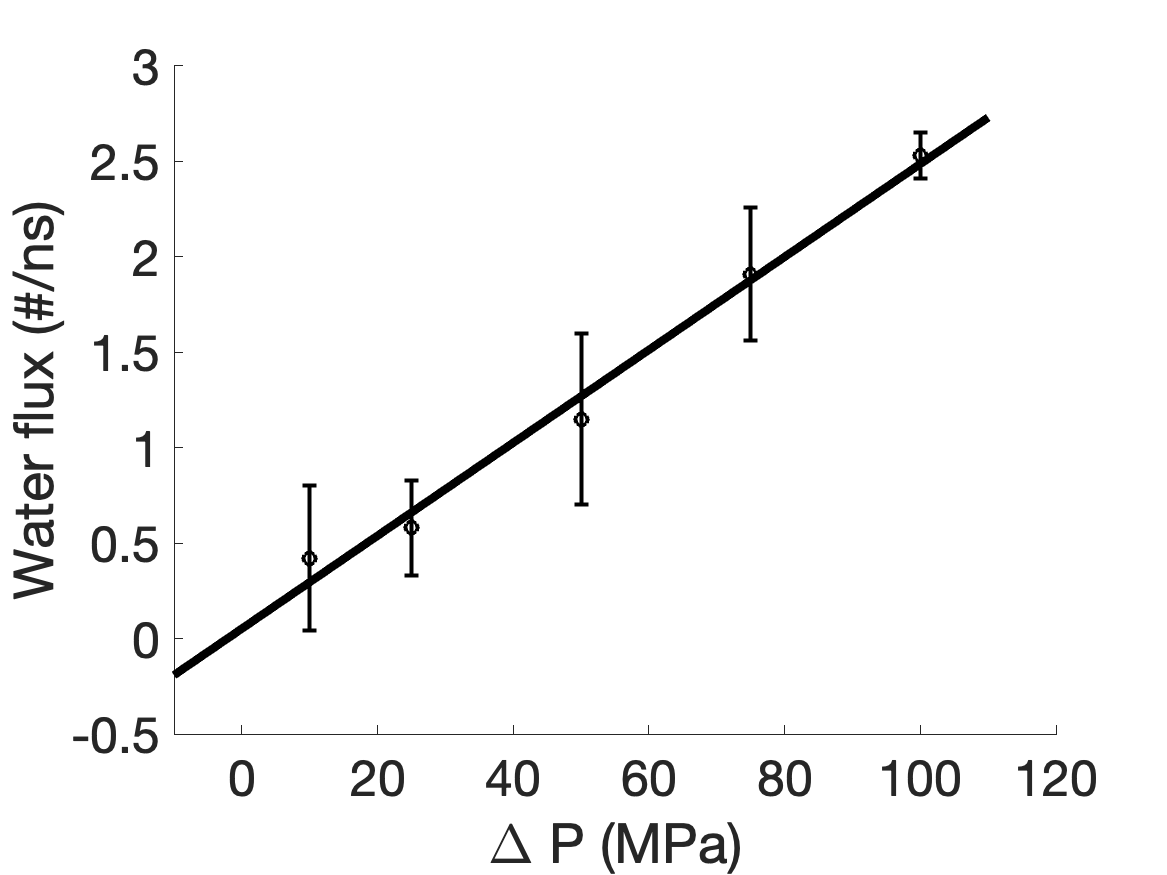}
 \caption{Dependence of the water flux on the 
 applied pressure difference. Error bars represent 
 the standard deviations of the water flux. 
 The line corresponding to the best-fit slope 
 for the five data points is shown.}
\label{fig7}
 \end{figure}

We consider five values for the hydrostatic 
pressure difference ranging from $10$ MPa to $100$ MPa, as shown 
in Fig.~\ref{fig7}. These values are sufficiently 
high to asses the measure of the water flux in 
a reasonable simulation time and sufficiently low 
to avoid the accumulation of water molecules in the 
vestibule of AqpZ, which could affect the linear relation between water flux and 
hydrostatic pressure difference~\cite{2002-BJ-Zhu-Schulten}. 
For comparaison, the osmotic pressure of physiological 
solutions is usually below $10$ MPa~\cite{1962-CC-Hendry} whereas the osmotic 
pressure measured for sea water is 
$\sim 26$ bars and the pressure 
traditionally used in reverse osmosis desalination 
of sea water is $\sim 60$ bars~\cite{2018-ESTL-Davenport-Elimelech}.

The evolution of the volume flux, $J_v$, across the active membrane as a function 
of the hydrostatic pressure difference, $\Delta P$,
is shown in Fig.~\ref{fig7}. 
Each NEMD simulation was run for $\sim 70$ ns and 
the hydraulic permeability, $L_P$, of the membrane 
is determined by interpolating the best-fit slope  through the data points.
Importantly, we observe that the linear 
relation between the water flux and the hydrostatic pressure difference, as given in Eq.~\ref{hydraulicPerm}, holds. 
We then measure the value for the hydraulic
permeability 
$L_P = 7.3 \pm 0.5 \times 10^{-18} \textrm{cm}^5~ \textrm{N}^{-1}~\textrm{s}^{-1}$. 
Applying Eq.~\ref{OsmoticHydraulic} with 
$T=300 K$ we obtain the osmotic permeability $P_f=1.1 \pm 0.1 \times 10^{-13} \textrm{cm}^3~\textrm{s}^{-1}$. 
The numerical result is in excellent agreement 
with the experimental estimate of the osmotic 
water permeability of the AqpZ monomer 
measured in biological conditions, 
%
$P^{\textrm{exp}}_f \sim 1.0 \times 10^{-13} \textrm{cm}^3 /s$~\cite{1999-JMB-Borgnia-Agre}.
Very importantly, this result confirms that the biological functionality of the AqpZ monomer is preserved after its directed insertion 
into the silica nanopore.
%
%

%

\section*{Conclusions}
The extensive numerical simulations discussed 
above allowed us to explore the design and 
stability of a distinctive class of biohybrid 
active membrane layer built from the directed 
insertion of aquaporin-incorporated lipid 
nanodiscs into a model silica nanopore.
We described in details the mechanisms at play 
during the insertion of the biological membrane 
into the solid-state nanopore. 
In particular, we showed that 
the preferential interaction between the 
hydrophobic chain of the POPC lipid bilayer and the 
alkyl group functionalizing the interior surface 
of the pore played an essential role both in 
the depletion of water molecules in the pore 
and the sealing properties of the system.
We numerically assessed the osmotic 
permeability of the biohybrid membrane and compared it 
with the experimental permeability
of the AqpZ monomer in the biological environment. 
In particular, we showed that the biological 
function of the AqpZ monomer, in terms of water transport, 
was preserved after the directed insertion 
into the silica nanopore. 
Experimental verification is required to test our 
expectations.\\

The biohybrid nanopore technology 
demonstrated in this work can be instrumental to 
the development of robust, thin, and defect-free 
membranes, with high selectivity and permeability.
In particular, incorporating aquaporin proteins 
into compatible solid-state nanopores, while ensuring 
industrial production scalability, will be decisive 
for translating the nano-scale science into 
 a large range of environmental 
and biomedical applications, including  
water desalination and aquaporin drug discovery.

 In the former case, membrane technology 
 has been dominating the water 
 desalination industry for decades due to 
 its high efficiency and 
 reliability~\cite{2018-FD-pedersen-Helix}. 
 However, the experimental performance of 
 biomimetic desalination membranes remains far 
 below that of biological 
 membranes~\cite{2020-ACSNano-Porter-Elimelech}.
The incorporation of aquaporin proteins into 
compatible solid-state nanopores provides 
biohybrid technology with the potential for 
site-specific genetic engineering or chemical 
modifications, which could potentially be used as active membrane layer  
in both reverse osmosis and forward osmosis 
applications to meet the increasing demand 
for fresh water at lower energy consumption 
and operating costs.

In the latter case, aquaporins are attractive 
targets for the development of novel drug therapies 
for life-threatening disorders, for example, 
kidney disease, congestive heart failure, stroke, 
traumatic injury and tumor-induced brain swelling, 
which currently lack adequate medical 
treatment~\cite{2005-DDT-Castle}.
Identifying new drug development frameworks for 
conditions associated with aberrant water movement 
will meet the urgent clinical need of millions of 
patients for whom no pharmacological interventions 
currently exist.
A key issue 
contributing to this situation may be the limited 
number of \textit{in vitro} assays suitable for 
high-throughput screening (HTS) of the pharmacological 
regulation of aquaporin function~\cite{2022-TPS-Salman-Bill}. 
In this perspective, the possibility of 
stabilizing aquaporin-incorporated lipid nanodiscs 
in compatible solid-state nanopore might be 
a promising way towards a reproducible mass production of robust membranes with a multitude 
of identical nanopores. This could 
pave the way to the design of transformative 
HTS drug discovery platforms to advance 
aquaporin-targeted therapeutics.

%


 
 
 \section*{Acknowledements}
This work was supported by the  UK Engineering and Physical Sciences Research Council (EPSRC) under grant number EP/V04804X/1. We are grateful to the UK Materials and Molecular Modelling Hub for computational resources, which 
 is partially funded by the EPSRC under grant numbers EP/P020194/1 
 and EP/T022213/1. FS thanks Kai Bin Yu, Turan Selman Erkal, and Fan Li for useful discussions. 
 
\normalem
\bibliographystyle{achemso.bst}
\bibliography{acs}
%
\pagebreak
\widetext
\begin{center}
\textbf{\large Biohybrid membrane formation by directed insertion of Aquaporin in to a solid-state nanopore} \end{center}

\begin{center}\textbf{\large Supporting Information}
\end{center}
\setcounter{equation}{0}
\setcounter{figure}{0}
\setcounter{table}{0}
\setcounter{page}{1}
\makeatletter
\renewcommand{\theequation}{S\arabic{equation}}
\renewcommand{\thefigure}{S\arabic{figure}}

\vskip 1.0cm
\section*{System characteristics}
\textbf{Aquaporin Z monomer.} The crystal structure of \textit{E. coli} AqpZ 
was obtained from the Protein Data Bank (PDB), 
entry 1RC2~\cite{2003-PB-Savage-Stroud}. 
The PDB file contains two protomers, from which 
protomer A was used to build the system simulated 
in this work. 
Missing side chains of residues Arg3, Glu31, Ser104, 
Arg230, and Asp231, as well as missing hydrogen atoms, 
were added using the programs Phyre2~\cite{2015-NP-Kelley-Sternberg} and VMD~\cite{1996-JMG-Humphrey-Schulten}. 
Titrable side chains were simulated in their default 
titration state, \textit{i.e.} Glu and Asp residues 
with a negative charge, Lys and Arg residues with a 
positive charge, and all other side chains with 
zero charge. Therefore, the protein 
is electrically neutral and no counterions were needed 
to ensure the electroneutrality of the AqpZ monomer. 
Finally, the program DiANNA~\cite{2005-NAR-Ferre-Clote} 
was used to check for the presence of any disulfide bond.\\

\textbf{AqpZ-incorporated lipid nanodisc.} The program \textit{Nanodisc Builder}~\cite{2019-CC-Qi-Im} in CHARMM-GUI~\cite{2008-CC-Jo-Im} 
was used to build the Aquaporin-incorporated lipid nanodisc. 
One AqpZ monomer was embedded in a 
$1$-palmitoyl-$2$-oleoyl-sn-glycero-$3$-phosphocholine 
(POPC) lipid-nanodisc stabilized by the MSP1E2D1 membrane 
scaffold protein. The diameter and net charge of the 
corresponding nanodisc are $11.1$ nm and $-14$e, respectively~\cite{2021-Membranes-Errasti-Palacin}. 
The number of POPC lipid molecules in the discoidal lipid 
bilayer is $230$. The net charge of the MSP nanodisc was 
neutralized with $14$ sodium ions.\\

\textbf{Silica nanopore.} The model solid-state nanopore was built 
from crystalline silica. The model of unit 
cell of cristobalite was obtained from the shared 
library of Materials Studio~\cite{MaterialsStudio}, 
which was periodically replicated in all directions. 
A regular silica slab was then carved out from the 
bulk cristobalite silica. 
Unsaturated silica atoms were
removed from the surface of the slab and the non-bridging oxygen atoms were saturated 
with hydroxyl (-OH) groups.
The structure was processed to saturate all valences, 
followed by energy minimization and molecular dynamics 
simulation to achieve structural stability. 
The volume of the equilibrated silica slab is 
$\sim (140 \times 140 \times 95)~\AA^3$. 
The resulting silica interface has a density of silanol 
groups of $6.4$ OH/$\textrm{nm}^2$, within the range 
of experimental data~\cite{2014-CM-Emami-Heinz}.
Next, a $90~\AA$ diameter cylindrical void was 
created in the center of the silica slab. 
%
Unsaturated silica atoms were removed from 
the interior surface of the pore and the non-bridging 
oxygen atoms were saturated with 
alkyl group $-\textrm{C}_n \textrm{H}_{2n+1}$, 
followed by energy minimization and molecular dynamics 
simulation to achieve structural stability. 
We considered different  functionalizing group 
with $n=1$, $2$, and $3$.
The ideal silica membrane model was adopted in this work to study the effect of interaction between the biological membrane and the pore surface on sealing property
%

\section*{Simulation details}
MD simulations were performed with the GROMACS software 
package, version 2018.8~\cite{2015-SoftwareX-Abraham-Lindahl}. 
Biased simulations were performed using version
2.3 of the plugin for free energy (FE) calculations, 
PLUMED~\cite{2014-CPC-Tribello-Bussi}.
The CHARMM force field~\cite{2016-NatureMethods-Huang-MacKerell} 
was employed to model the biological entities (MSP, AqpZ, POPC) and the alkyl groups on the silica surface. 
The INTERFACE force field~\cite{2013-Langmuir-Heinz-Emami} 
and TIP3P potential~\cite{1983-JChemPhys-Jorgensen-Madura} 
were employed to model the silica nanopore and water molecules, 
respectively, as they are compatible with the CHARMM force field.
The time step used in all the simulations is $0.002$ ps, 
and the list of neighbors is updated every $0.04$ ps with 
the grid method and a cutoff radius of $1.2$ nm.
The LINear Constraint Solver (LINCS) algorithm~\cite{1997-JCC-Hess-Fraaije} 
handled bond constraints while the particle-mesh Ewald 
scheme~\cite{1993-JCP-Darden-Pedersen} was used to treat long-range electrostatic interactions. The nonbonded van der Waals cutoff radius 
was $1.2$ nm. The initial velocities were chosen randomly. 
The cutoff algorithm was applied for the non-Coulomb 
potentials with a radius of $1.2$ nm.\\

The energy of the system was first optimized using 
the steepest descent minimization algorithm.
The system was then equilibrated for $250$ ps 
at constant temperature, $T=300$ K, using the Berendsen 
thermostat with lower restraints. 
The equilibration phase was then conducted within the 
isobaric-isothermal (NPT) ensemble to equilibrate the 
fluid density. The temperature and pressure were maintained 
at $T=300$ K and $P=1$ bar, respectively, using the 
Berendsen thermostat and barostat for $2$ ns. 
We then switched to the Nose-Hoover thermostat 
and the Parrinello-Rahman barostat for an additional $2$ ns, 
which are considered more thermodynamically consistent 
algorithms.
Finally the simulation was continued 
in NVT conditions coupling with the v-rescale thermostat 
at constant temperature $T=300$ K.\\

\begin{figure}[b]
\includegraphics[width=0.55 \textwidth, angle=-0]{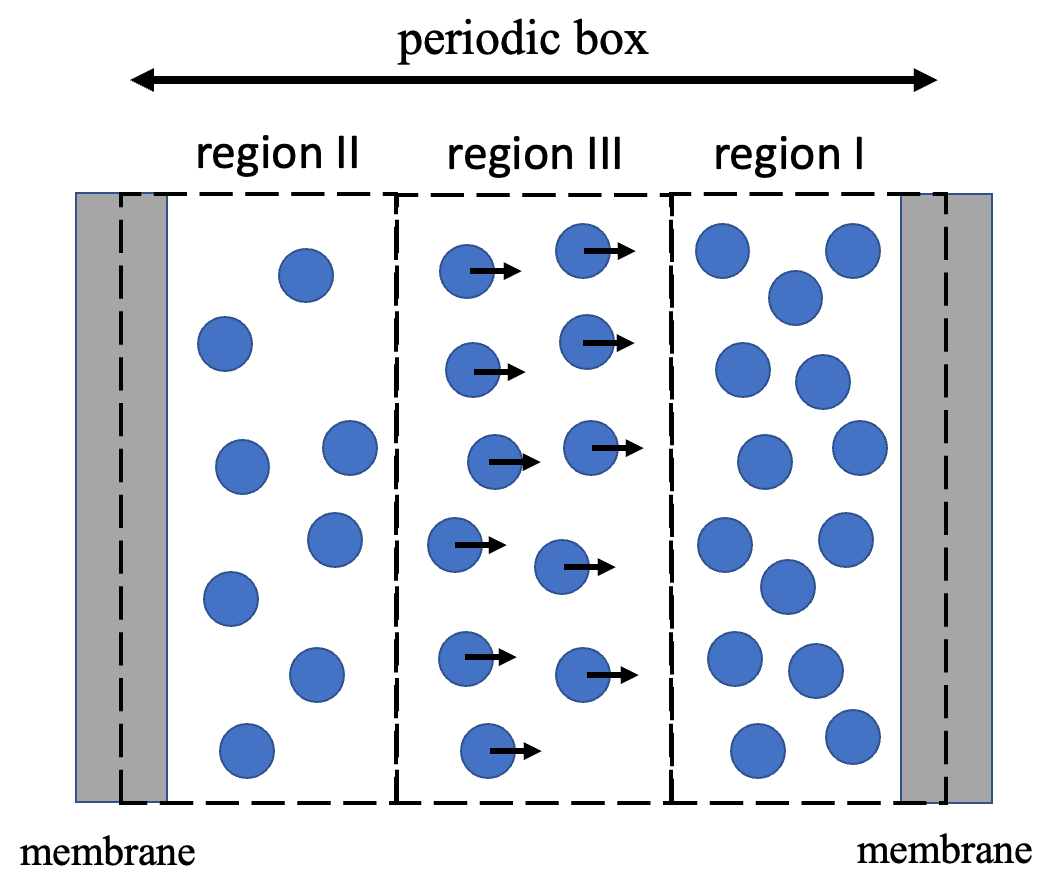}
 \caption{Schematic representation of the numerical 
 method implemented to produce a pressure difference in
MD simulations using periodic boundary conditions. 
A constant force $f$ is exerted on the oxygen atoms of the water molecules 
in region III. The membranes shown in the figure 
are the periodic images of each other. }
\label{fig1SI}
 \end{figure}
To achieve the insertion of the Aqp-incorporated lipid nanodisc 
into the solid-state nanopore and to measure the osmotic permeability 
of the system, a hydrostatic pressure difference was generated 
across the membrane using the method proposed by 
Zhu \textit{et al.}~\cite{2002-BJ-Zhu-Schulten,2004-BJ-Zhu-Schulten} and the extension of the pull module 
of the GROMACS software developed by Gr{\"a}ter 
and coworkers~\cite{2019-BJ-Herrera-Grater}. 
As shown in Fig.~\ref{fig1SI}, three regions (I, II, III)  were 
defined in the water layer with region III isolated from 
the two sides of the membrane by region I and II, respectively. 
A constant force $f$ along the $z$-direction is exerted on 
all water molecules in region III, generating a pressure 
gradient in this region. This results in a pressure difference 
between regions I and II
\begin{equation}
    \Delta P = P_1 - P_2 = n f/A \,,\nonumber
\end{equation}
where $n$ is the number of water molecules in region III, 
and $A$ is the area of the membrane. To prevent the overall 
translation of the whole system along the direction 
of the applied forces, position constraints were applied on 
the membrane in the $z$-direction~\cite{2004-BJ-Zhu-Schulten}. 
In particular, we applied the  constant force $(f)$ on the oxygen atoms of 
the water molecules in region III, defined as a $8$ nm thick 
layer.
We applied harmonic constraints to the $C_{\alpha}$ atoms 
of the proteins, the phosphorus atoms 
of the lipid molecules, and the silicon and oxygen atoms of 
the silica pore, with spring constants of 
$50~\textrm{kJ}/\textrm{mol}/\textrm{nm}^2$, 
$700~\textrm{kJ}/\textrm{mol}/\textrm{nm}^2$, 
and $700~\textrm{kJ}/\textrm{mol}/\textrm{nm}^2$, 
respectively.

\section*{Sealing properties of the biohybrid nanopore}
\begin{figure}[h]
\includegraphics[width=0.5 \textwidth, angle=-0]{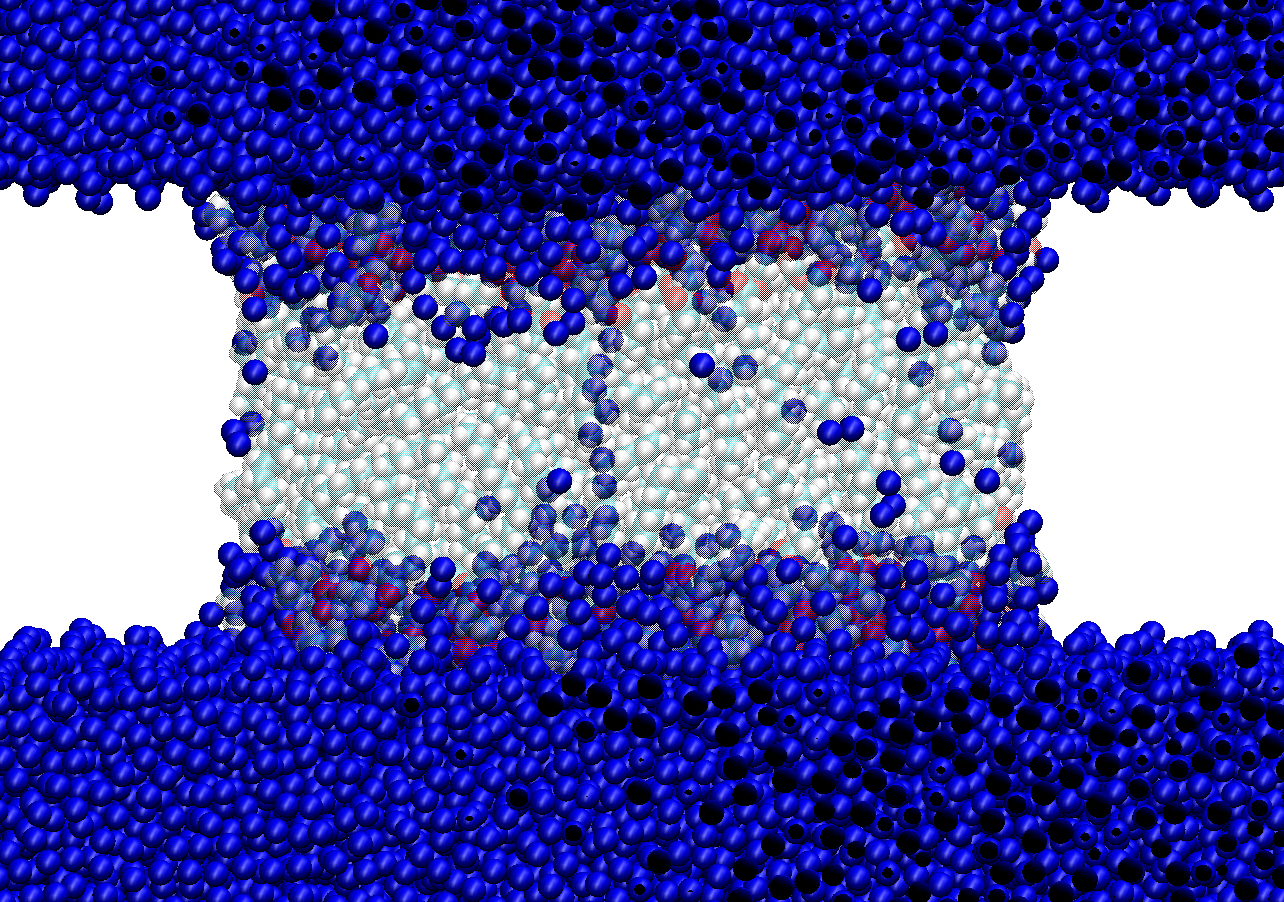}
 \caption{Schematic representation of the 
 Aqp-incorporated lipid shell inserted in the 
 silica nanopore functionalized with the butyl group, 
 after the system relaxed for $1$ ns. 
 Red, blue, white, gold, and cyan spheres 
 represent oxygen, nitrogen, hydrogen, 
 phosphate and carbon atoms in the lipid molecules, 
 respectively. The oxygen atoms of the water molecules 
 are shown in blue.
 The lipid molecules are shadowed 
 and the silica nanopore and the AqpZ monomer 
 are not shown for clarity.
 In addition to the water channel which connects 
 both sides of the membrane, we observe the presence 
 of water molecules at the interface between the lipid molecules 
 and the interior surface of the silica pore.}
\label{fig2SI}
 \end{figure}

We studied the effect of the length of the alkyl group 
$-\textrm{C}_n \textrm{H}_{2n+1}$, which functionalizes the 
interior surface of the silica pore, on the sealing property 
of the biohybrid membrane. In particular, we considered 
methyl ($n=1$), butyl ($n=2$), and propyl ($n=3$) groups. 
In Fig.~\ref{fig2SI} is shown the schematic representation of 
the system after it relaxed for $1$ ns, when the interior 
surface of the pore was functionalized with the butyl group. 
We observed the presence of water molecules at 
the interface between the lipid molecules and the 
interior surface of the silica pore, which can be 
explained by the abscence of preferential interaction between the hydrophobic tail of the lipids and the butyl groups.

 \section*{Density  distribution of ions in sea water environment}
\begin{figure}[t]
\includegraphics[width=0.8 \textwidth, angle=-0]{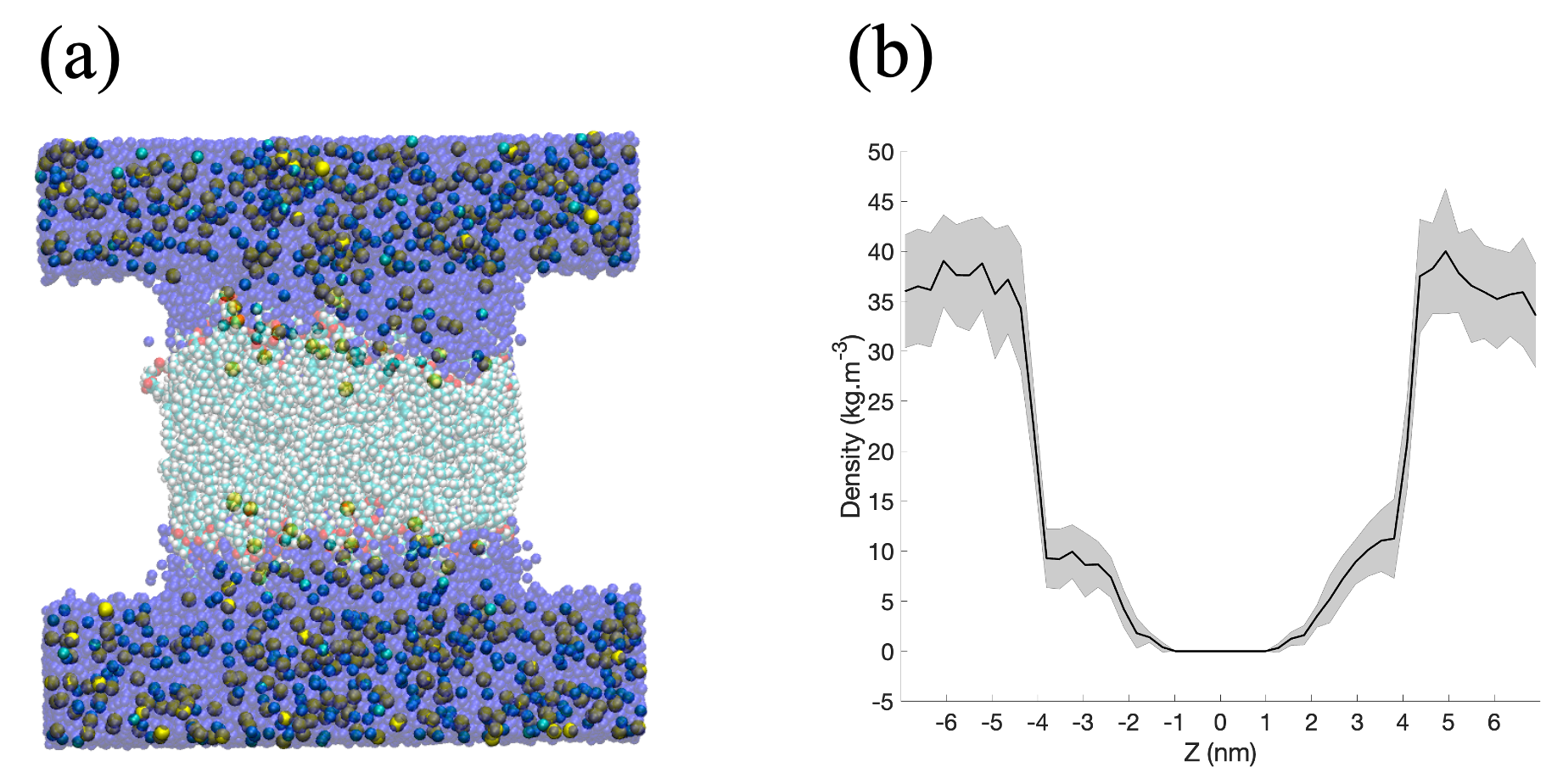}
 \caption{(a) Schematic representation of the system 
 in sea saline water. The solid-state nanopore is not shown 
 for clarity. The chloride and sodium ions are shown in green 
 and yellow, respectively. Red, blue, white, gold, and cyan spheres 
 represent oxygen, nitrogen, hydrogen, 
 phosphate and carbon atoms in the lipid molecules, respectively. 
 The oxygen atoms of the water molecules 
 are shown in blue.
 (b) Density profile of ions (sodium and chloride ions) 
 along the Z direction of the 
 simulation box centered around the 
 ApqZ monomer, measured after the system relaxed 
 for $200$ ns in sea water environment. 
 Uncertainties, defined as the standard errors, are represented by the shaded 
area.}
\label{fig3SI}
 \end{figure}

In Fig.~\ref{fig3SI} is shown the density distribution of ions 
(sodium and chloride ions) in the simulation 
box after the system relaxed for $200$ ns at constant 
temperature $T=300$ K.
The ion density in the Aqp-incorporated lipid shell, 
i.e. $\vert z \vert \leq 1$,  is exactly zero, indicating 
the efficiency of the membrane to play the role 
of separator. 
For $1 \leq \vert z \vert \leq 4$ we observe 
the progressive increase of the ion density, 
similar to the one observed for the water molecules 
(cf. main text). 
This profile is explained by the conformational 
organisation of the lipid bilayer in the nanopore 
due to preferential interaction with the hydrophobic 
surface.
For $\vert z \vert \geq 4$, \textit{i.e.} outside 
the solid-state nanopore, the system transitions 
abruptly to reach the bulk desired ion density.
 
 \section*{Equilibrium dissociation of 
 the AqpZ monomer from the MSP-based lipid 
 nanodisc}
\begin{figure}[h]
\includegraphics[width=0.5 \textwidth, angle=-0]{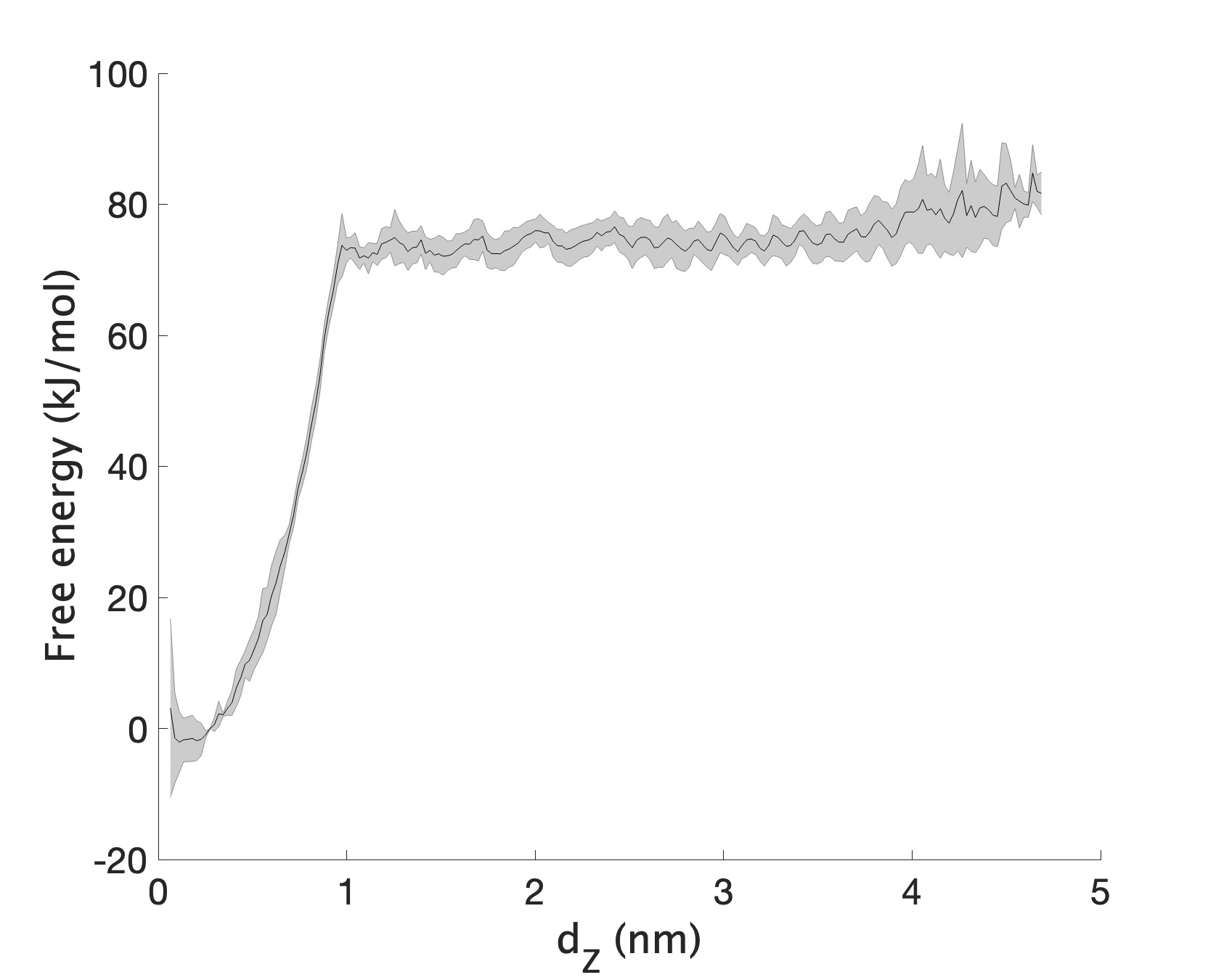}
 \caption{FEP associated with the 
 equilibrium dissociation of the AqpZ monomer from the 
 MSP-based lipid nanodisc, obtained within the US/ABMD framework 
 and calculated with DHAM and a number of bin of 200. 
 The $x$-axis corresponds to the projection of the Cartesian 
 distance between the center of mass of AqpZ and the center of mass 
 of the MSPs, $\textrm{d}_Z$.  The difference in FE between the global minima
($\textrm{d}_Z \approx 0.1$)  and the region where the FEP 
flattens ($\textrm{d}_Z \approx 1$) is $\Delta F_0 \approx 75~\textrm{kJ/mol}$. Uncertainties, defined as the standard errors, are represented by the shaded 
area for US data.}
\label{fig4SI}
 \end{figure}

 The phenomenon of interest (\textit{i.e.} the dissociation 
 of the AqpZ monomer from the MSP-based lipid 
 nanodisc) occurs on time scales that are orders of magnitude 
longer than those accessible with classical 
MD simulations. 
To overcome this limitation, we use the numerical method 
based on constrained MD employed by Sicard 
and coworkers,~\cite{2018-Langmuir-Sicard-Striolo,2019-ACSNano-Sicard-Striolo} 
which combined the adiabatic biased molecular dynamics (ABMD)~\cite{1999-JMB-Paci-Karplus,1999-JCP-Marchi-Ballone, 2011-JCP-Camilloni-Tiana,2016-FD-Sicard-Striolo} 
and umbrella sampling (US)~\cite{2011-CMS-Kastner} frameworks. 
To design the US windows, we use the projection of 
the Cartesian distance between the center of mass of 
the AqpZ monomer and the center of mass of the MSPs along the $Z$-direction, $\textrm{d}_Z$, as a reaction coordinate (RC). 
The starting configurations for the US simulations are obtained 
by pulling adiabatically the system along the RC, 
generating $180$ windows. 
Each US window is subsequently run for $2$ ns to allow equilibration, 
followed by additional $2$ ns of the production run.
Upon completion of the US simulations, we obtain the free energy 
profile (FEP) associated with the dissociation of the 
AqpZ monomer using the dynamic histogram analysis method 
(DHAM).~\cite{2015-JCTC-Rosta-Hummer} 
This unbiasing method uses a maximum likelihood estimate 
of the Markov State Model (MSM) transition probabilities given 
the observed transition counts during each biased trajectory. 
To produce MSMs 
from the biased simulations, we discretize the RC into bins 
and count the number of transitions between each pair of bins 
$i$ and $j$ 
in the US simulation $k$ at the chosen lagtime $t$, $C_{ij}^k(t)$, 
as well as the number of times each bin is occupied during each 
US simulation $k$, $n_i^k = \sum_j C_{ij}^k(t)$. 
These values provide the conditional probabilities 
$M_{ij}(t)=P(j,t \vert i,0)$. For biased simulations, 
where a biasing energy of $u_i^k$ is applied to state $i$ during simulation $k$, 
we compute the unbiased MSM from the biased data, as given by 
\begin{equation}
M_{ij}(t) = \frac{\sum_k C_{ij}^k(t)}{\sum_k n_i^k~\exp\big( -(u_j^k-u_i^k)/2k_BT\big)}. \nonumber
\label{DHAMeq}
\end{equation}
Once the MSM is constructed from the simulation data, the equilibrium 
probabilities are calculated as the eigenvector corresponding to eigenvalue 
$1$ of the transition matrix $M_{ij}$.
In Fig.~\ref{fig4SI} is shown the FEP associated with 
the dissocition of the AqpZ monomer 
plotted along the projection of the Cartesian distance between 
the center of mass of AqpZ and the center of mass of the MSPs. 
Uncertainties were determined by dividing the data into 
three equal sections, determining the profiles independently, 
and calculating the standard error. 
We measured a difference in FE between the global minimum
($\textrm{d}_Z \approx 0.1$)  and the region where the FEP 
flattens ($\textrm{d}_Z \approx 1$), $\Delta F_0 \approx 75~\textrm{kJ/mol}$.

\end{document}